\documentclass[aps,prl, preprintnumbers,groupedaddress,nofootinbib,twocolumn]{revtex4}

\usepackage[dvipdfmx]{graphicx}
\usepackage{amssymb,amsmath}

\begin{document}

\begin{titlepage}
\begin{flushright}
\begin{tabular}{l}
 SU-HET-10-2015\\
\end{tabular} 
\end{flushright}
\end{titlepage}

\title{Bosonic seesaw mechanism in a classically conformal extension of the Standard Model 
}

\author{Naoyuki Haba$^1$, Hiroyuki Ishida$^1$, Nobuchika Okada$^2$, and Yuya Yamaguchi$^{1,3}$\vspace{2mm}}
\affiliation{$^1$ Graduate School of Science and Engineering, Shimane University, Matsue 690-8504, Japan\\
$^2$ Department of Physics and Astronomy, University of Alabama, Tuscaloosa, Alabama 35487, USA\\
$^3$ Department of Physics, Faculty of Science, Hokkaido University, Sapporo 060-0810, Japan}

\begin{abstract}
We suggest the so-called bosonic seesaw mechanism in the context of a classically conformal 
   $U(1)_{B-L}$ extension of the Standard Model with two Higgs doublet fields. 
The $U(1)_{B-L}$ symmetry is radiatively broken via the Coleman-Weinberg mechanism, 
   which also generates the mass terms for the two Higgs doublets through quartic Higgs couplings. 
Their masses are all positive but, nevertheless, the electroweak symmetry breaking 
   is realized by the bosonic seesaw mechanism. 
Analyzing the renormalization group evolutions for all model couplings, 
   we find that a large hierarchy among the quartic Higgs couplings, 
   which is crucial for the bosonic seesaw mechanism to work, 
   is dramatically reduced toward high energies. 
Therefore, the bosonic seesaw is naturally realized with only a mild hierarchy, 
   if some fundamental theory, which provides the origin of the classically conformal invariance, 
   completes our model at some high energy, for example, the Planck scale.
We identify the regions of model parameters which satisfy the perturbativity of the running couplings
 and the electroweak vacuum stability as well as the naturalness of the electroweak scale.  
\end{abstract}

\maketitle


In the Standard Model (SM),
  the electroweak symmetry breaking is realized by the negative mass term in the Higgs potential,
  which seems to be artificial because there is nothing to stabilize the electroweak scale.
If new physics takes place at a very high energy, e.g. the Planck scale,
  the mass term receives large corrections which are quadratically sensitive to the new physics scale, 
  so that the electroweak scale is not stable against the corrections.  
This is the so-called gauge hierarchy problem.
It is well known that supersymmetry (SUSY) can solve this problem. 
Since the mass corrections are completely canceled by the SUSY partners,
  no fine-tuning is necessary to reproduce the electroweak scale correctly, 
  unless the SUSY breaking scale is much higher than the electroweak scale.
On the other hand, since no indication of SUSY particles has been obtained in the large hadron collider (LHC) experiments, 
  one may consider other solutions to the gauge hierarchy problem without SUSY.

In this direction, recently a lot of works have been done in models based on a classically conformal symmetry.
There are $U(1)$ gauge extension~\cite{Hempfling:1996ht}-\cite{Plascencia:2015xwa},
 and non-Abelian gauge extension, in which conformal symmetry is broken by
 radiative corrections~\cite{Khoze:2014xha,Holthausen:2009uc,Heikinheimo:2013fta,Heikinheimo:2013xua,Hambye:2013sna,Carone:2013wla}
 and strong dynamics~\cite{Hur:2011sv}-\cite{Kubo:2015joa}.
In addition, there are also non-gauge extended models [see Ref.~\cite{Hashino:2015nxa} and therein].\footnote{
In Ref.~\cite{Hashino:2015nxa},
 the upper bound on the mass of the lightest additional scalar boson is obtained as $\simeq 543\,{\rm GeV}$,
 which is independent of its isospin and hypercharge.
Thus, the classically conformal model is strongly constrained without gauge extension.}
This direction is based on the argument by Bardeen~\cite{Bardeen:1995kv} that 
  the quadratic divergence in the Higgs mass corrections can be subtracted 
  by a boundary condition of some ultraviolet complete theory, 
  which is classically conformal, and only logarithmic divergences should be considered
  (see Ref.~\cite{Iso:2012jn} for more detailed discussions). 
If this is the case, imposing the classically conformal symmetry to the theory 
  is another way to solve the gauge hierarchy problem. 
Since there is no dimensionful parameter in this class of models,
   the classically conformal symmetry must be broken by quantum corrections. 
This structure fits the model first proposed by Coleman and Weinberg~\cite{Coleman:1973jx}, 
   where a model is defined as a massless theory and the classically conformal symmetry is radiatively broken 
   by the Coleman-Weinberg (CW) mechanism, generating  a mass scale through the dimensional transmutation.

In this paper we propose a classically conformal $U(1)_{B-L}$ extended SM with two Higgs doublets.  
An SM singlet, $B-L$ Higgs field develops its vacuum expectation value (VEV) by the CW mechanism, 
  and the $U(1)_{B-L}$ symmetry is radiatively broken. 
This gauge symmetry breaking also generates the mass terms for the two Higgs doublets 
  through quartic couplings between the two Higgs doublets and the $B-L$ Higgs field. 
We assume the quartic couplings to be all positive at the $U(1)_{B-L}$ breaking scale
  but, nevertheless, the electroweak symmetry breaking 
  is triggered through the so-called bosonic seesaw mechanism~\cite{Calmet:2002rf,Kim:2005qb,Haba:2005jq}, 
  which is analogous to the seesaw mechanism for the neutrino mass generation 
  and leads to a negative mass squared for the SM-like Higgs doublet.  
Because a negative quartic coupling may cause vacuum instability,
 it is important to take all quartic couplings to be positive,
 while in the conventional models, e.g., Refs~\cite{Iso:2009ss} and \cite{Hur:2011sv},
 the mixing coupling between the $SU(2)_L$ doublet and singlet fields is necessarily negative
 to realize the negative mass term of the SM-like Higgs doublet.
Our model guarantees that the mixing couplings are positive at the breaking scale
 with a hierarchy among the quartic couplings,
 which successfully derives the bosonic seesaw mechanism.
The hierarchy seems to be unnatural,
 but we find that the renormalization group evolutions of the quartic couplings 
 dramatically reduce the large hierarchy toward high energies. 
On the other hand,
 a large hierarchy exists even in the conventional model,
 that is, the mixing coupling should be much small as $({\rm EW\ scale})^2/v^2$
 with a conformal symmetry breaking scale $v$,
 except for $v \sim {\cal O}(1)\,{\rm TeV}$.
Note that the degree of the hierarchy in our model does not increase
 as the symmetry breaking scale becomes larger.

In the following, let us explain our model in detail.
We consider an extension of the SM with an additional $U(1)_{B-L}$ gauge symmetry. 
Our model has three scalar fields,
 that is, two Higgs doublets ($H_1$ and  $H_2$) and one SM singlet, $B-L$ Higgs field ($\Phi$) are introduced.
The $U(1)_{B-L}$ charges of $H_1$, $H_2$, and $\Phi$ are 0, 4, and 2, respectively. 
As is well known, the introduction of the three right-handed neutrinos ($N^i$, $i=1$, 2, 3)
 with a $U(1)_{B-L}$ charge
 is crucial to make the model free from all the gauge and gravitational anomalies. 
In addition, we impose a classically conformal symmetry to the model, 
  under which the scalar potential is given by
\begin{eqnarray}
	V &=& \lambda_1 |H_1|^4 + \lambda_2 |H_2|^4 + \lambda_3 |H_1|^2 |H_2|^2 \nonumber \\
	&& + \lambda_4 (H_2^\dagger H_1) (H_1^\dagger H_2) + \lambda_\Phi |\Phi|^4 
		+ \lambda_{H1\Phi} |H_1|^2 |\Phi|^2 \nonumber \\
	&&+ \lambda_{H2\Phi} |H_2|^2 |\Phi|^2 + \left( \lambda_{\rm mix} (H_2^\dagger H_1) \Phi^2 + h.c. \right).
\end{eqnarray}
Here, all of the dimensionful parameters are prohibited by the classically conformal symmetry.
In this system, the $U(1)_{B-L}$ symmetry must be radiatively broken by quantum effects, i.e., the CW mechanism.
The CW potential for $\Phi$ is described as
\begin{eqnarray}
	V_\Phi(\phi) = \frac{1}{4} \lambda_\Phi(v_\Phi)\, \phi^4
						+ \frac{1}{8} \beta_{\lambda_{\Phi}}(v_\Phi)\, \phi^4
							\left( \ln \frac{\phi^2}{v_\Phi^2} -\frac{25}{6} \right),
\label{CWpotential}
\end{eqnarray}
 where $\Re[\Phi] = \phi/\sqrt{2}$, and $v_\Phi=\langle \phi \rangle$ is the VEV of $\Phi$.
When the beta function $\beta_{\lambda_\Phi}$ is dominated by
  the $U(1)_{B-L}$ gauge coupling ($g_{B-L}$) and the Majorana Yukawa couplings of right-handed neutrinos ($Y_M$) ,
 the minimization condition of $V_\Phi$ approximately leads to
\begin{eqnarray}
	\lambda_\Phi \simeq \frac{11}{6 \pi^2}
		\left( 6 g_{B-L}^4 - {\rm tr} Y_M^4 \right),
\label{CW_relation}
\end{eqnarray}
  where all parameters are evaluated at $v_\Phi$.  
Through the $U(1)_{B-L}$ symmetry breaking, the mass terms of the two Higgs doublets arise 
  from the mixing terms between $H_{1,2}$ and $\Phi$, and the scalar mass squared matrix is read as 
\begin{eqnarray}
 	-\mathcal{L} &=& \frac{1}{2}
			(H_1, H_2) \left( \begin{array}{cc}
				\lambda_{H1\Phi} v_\Phi^2 & \lambda_{\rm mix} v_\Phi^2 \\
				\lambda_{\rm mix} v_\Phi^2 & \lambda_{H2\Phi} v_\Phi^2
			\end{array} \right)
	\left( \begin{array}{c} H_1 \\ H_2 \end{array} \right) \nonumber\\
			&\approx& \frac{1}{2}
			(H'_1, H'_2) \left( \begin{array}{cc}
				\lambda_{H1\Phi}v_\Phi^2 - \frac{\lambda_{\rm mix}^2 v_\Phi^2}{\lambda_{H2\Phi}} & 0 \\
				0 & \lambda_{H2\Phi} v_\Phi^2
			\end{array} \right)
	\left( \begin{array}{c} H'_1 \\ H'_2 \end{array} \right),\nonumber\\
\label{matrix}
\end{eqnarray}
 where $H'_1$ and $H'_2$ are mass eigenstates,
 and we have assumed a hierarchy among the quartic couplings as
 $0 \leq  \lambda_{H1\Phi} \ll \lambda_{\rm mix} \ll \lambda_{H2\Phi}$ at the scale $\mu = v_\Phi$.
In the next section, we will show that this hierarchy is dramatically reduced toward high energies 
  in their renormalization group evolutions. 
Because of this hierarchy,
  mass eigenstates $H'_1$ and $H'_2$ are almost composed of $H_1$ and $H_2$, respectively.
Hence, we approximately identify $H'_1$ with the SM-like Higgs doublet. 
Note that even though all quartic couplings are positive, 
  the SM-like Higgs doublet obtains a negative mass squared for $\lambda_{H1\Phi} \ll \lambda_{\rm mix}^2/\lambda_{H2\Phi}$, 
  and hence the electroweak symmetry is broken. 
This is the so-called bosonic seesaw mechanism \cite{Calmet:2002rf,Kim:2005qb, Haba:2005jq}.

In more precise analysis for the electroweak symmetry breaking, we take into account 
   a scalar one-loop diagram through the quartic couplings, $\lambda_3$ and $\lambda_4$, 
%
%
%
 and the SM-like Higgs doublet mass is given by
\begin{eqnarray}
	m_h^2 
			&\simeq& \lambda_{H2\Phi} v_\Phi^2 \left[ 
				\frac{1}{2} \left(\frac{\lambda_{\rm mix}}{\lambda_{H2\Phi}}\right)^2
				+ \frac{2\lambda_3 + \lambda_4}{16\pi^2}  \right],
\label{Higgs}
\end{eqnarray}
  where we have omitted the $\lambda_{H1\Phi}$ term in the second line, and 
  the observed Higgs boson mass $M_h=125$ GeV is given by $M_h=m_h/\sqrt{2}$.

In addition to the scalar one-loop diagram, 
  one may consider other Higgs mass corrections coming from a neutrino one-loop diagram and 
  two-loop diagrams involving the $U(1)_{B-L}$ gauge boson ($Z'$) and the top Yukawa coupling, 
  which are, respectively, found to be~\cite{Iso:2009ss}
\begin{eqnarray}
		\delta m_h^2 \sim \frac{Y_\nu^2 Y_M^2 v_\Phi^2}{16 \pi^2},\qquad
		\delta m_h^2 \sim \frac{y_t^2 g_{B-L}^4 v_\Phi^2}{(16 \pi^2)^2},
\label{deltam}
\end{eqnarray}
  where $Y_\nu$ and $y_t$ are Dirac Yukawa couplings of neutrino and top quark, respectively.  
It turns out that these contributions are negligibly small
 compared to the scalar one-loop correction in Eq.~(\ref{Higgs}). 
As we will discuss in the next section, the quartic couplings $\lambda_3$ and $\lambda_4$ 
  should be sizable $\lambda_{3,4} \gtrsim 0.15$ in order to stabilize the electroweak vacuum. 
The neutrino one-loop correction is roughly proportional to the active neutrino mass 
  by using the seesaw relation,  and it is highly suppressed by the lightness of the neutrino mass.
The two-loop corrections with the $Z'$ boson is suppressed by a two-loop factor $1/(16\pi^2)^2$. 
Unless $g_{B-L}$ is large, the two-loop corrections are smaller than the scalar one-loop correction. 


The other scalar masses are approximately given by
\begin{eqnarray}
	M_\phi^2 &=& \frac{6}{11} \lambda_\Phi v_\Phi^2, \\
	M_H^2 &=& M_A^2 = \lambda_{H2\Phi} v_\Phi^2 + (\lambda_3 + \lambda_4) v_H^2, \\
	M_{H^\pm}^2 &=& \lambda_{H2\Phi} v_\Phi^2 + \lambda_3 v_H^2, 
\end{eqnarray}
 where  $M_\phi$ is the mass of the SM singlet scalar,
 $M_H$ ($M_A$) is the mass of CP-even (CP-odd) neutral Higgs boson,
 and $M_{H\pm}$ is the mass of charged Higgs boson. 
The extra heavy Higgs bosons are almost degenerate in mass.   
The masses of the $Z'$ boson and the right-handed neutrinos are given by
\begin{eqnarray}
	M_{Z'} &=& 2 g_{B-L} v_\Phi,
\label{MZ} \\
	M_N &=& \sqrt{2} y_M v_\Phi
		\simeq \left[ \frac{3}{2N_\nu} \left( 1 - \frac{\pi^2 \lambda_\Phi}{11 g_{B-L}^4} \right) \right]^{1/4} M_{Z'},
\label{MN}
\end{eqnarray}
 where we have used ${\rm tr}Y_M=N_\nu y_M$, for simplicity,
 and $N_\nu$ stands for the number of relevant Majorana couplings.
In the following analysis, we will take $N_\nu = 1$ for simplicity,
  because our final results are almost insensitive to $N_\nu$. 
In the last equality in Eq.~(\ref{MN}), we have used Eq.~(\ref{CW_relation}).


Before presenting our numerical results,
   we first discuss constraints on the model parameters from the perturbativity  
   and the stability of the electroweak vacuum in the renormalization group evolutions. 
In our analysis, all values of couplings are given at $\mu=v_\Phi$.
For $v_\Phi$ at the TeV scale, 
   we find the constraint $g_{B-L}\lesssim 0.3$ to avoid the Landau pole of the gauge coupling 
   below the Planck scale, while a more severe constraint $g_{B-L} \lesssim 0.2$ is obtained 
   to avoid a blowup of  the quartic coupling $\lambda_2$ below the Planck scale. 
From $g_{B-L} \lesssim 0.2$ and
 the experimental bound $M_{Z'} > 2.9$ TeV on the $Z'$ boson mass \cite{Aad:2014cka,Khachatryan:2014fba},
 we find $v_\Phi > 7.25$\,TeV.
The electroweak vacuum stability, in other words, $\lambda_H(\mu) > 0$ for any scales between  
  the electroweak scale and the Planck scale, 
  can be realized by sufficiently large $\lambda_3$ and/or $\lambda_4$
  as $\lambda_3 =\lambda_4 \gtrsim 0.15$.
To keep their perturbativity below the Planck scale, 
 $\lambda_3 =\lambda_4 \lesssim 0.48$ must be satisfied,
 while we will find that the naturalness of the electroweak scale leads to a more severe upper bound.


To realize the hierarchy $\lambda_{H1\Phi} \ll \lambda_{\rm mix} \ll \lambda_{H2\Phi}$,
  we take $\lambda_{H1\Phi}=0$, for simplicity. 
When we consider $\lambda_{\rm mix}$ in the range of $0< \lambda_{\rm mix} < 0.1\times \lambda_{H2\Phi}$,
   the relation between $v_\Phi$ and $\lambda_{H2\Phi}$ obtained by Eq.~(\ref{Higgs})
 is almost uniquely determined.
When we fix $\lambda_3 =\lambda_4 = 0.15$ as an example,
 we find $1\,{\rm TeV} \lesssim \lambda_{H2\Phi} v_\Phi^2 \lesssim 1.7\,{\rm TeV}$
 for $v_\Phi \gtrsim 10$\,TeV, which is almost independent of $v_\Phi$. 
Since all heavy Higgs boson masses are approximately determined by $\lambda_{H2\Phi} v_\Phi^2$,
  they lie in the range between 1\,TeV and 1.7\,TeV.
Such heavy Higgs bosons can be tested at the LHC in the near future. 

In Eq.~(\ref{Higgs}), it may be natural for the first term from the tree-level couplings 
   dominates over the second term from the 1-loop correction. 
This naturalness leads to the constraint of  $\lambda_3 = \lambda_4 < 0.26$,
   which is more severe than the perturbativity bound $\lambda_3 =\lambda_4 \lesssim 0.48$ discussed above.
This condition is equivalent to the fact that the origin of the negative mass term mainly comes from
  the diagonalization of the scalar mass squared matrix in Eq.~(\ref{matrix}),
  namely, the bosonic seesaw mechanism.

\begin{figure}[t]
\begin{center}
          \includegraphics[clip, height=6cm]{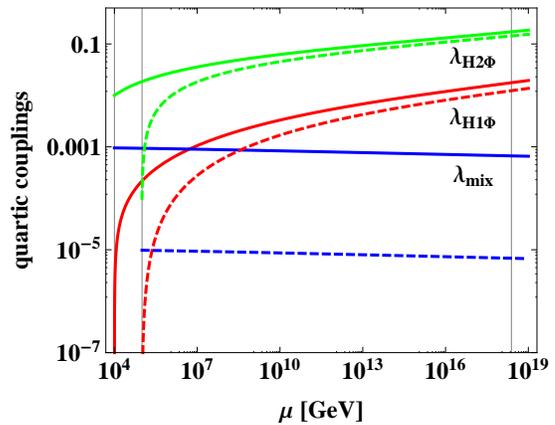} 
\end{center}
\caption{
Renormalization group evolutions of the quartic couplings for $v_\Phi=10$\,TeV (solid) and 100\,TeV (dashed).
The red, green, and blue lines correspond to
 $\lambda_{H1\Phi}$, $\lambda_{H2\Phi}$ and $\lambda_{\rm mix}$, respectively.
The rightmost vertical line shows the reduced Planck scale.
}
\label{running}
\end{figure}

Now we present the results of our numerical analysis. 
In Fig.~\ref{running}, we show the renormalization group evolutions of the quartic couplings.
Here, we have taken $\lambda_{H1\Phi} = 0$,
 and $\lambda_{H2\Phi} = 10^{-2}$ and $10^{-4}$
 for $v_\Phi=10$\,TeV (solid lines) and 100\,TeV (dashed lines), respectively.
The red, green, and blue lines correspond to the running of
 $\lambda_{H1\Phi}$, $\lambda_{H2\Phi}$ and $\lambda_{\rm mix}$, respectively.
The rightmost vertical line denotes the reduced Planck scale $M_{Pl} = 2.4 \times 10^{18}$ GeV.
In this plot, the other input parameters have been set
  as $g_{B-L}  = 0.17$ and $\lambda_3  = \lambda_4  = 0.17$
  to realize the electroweak vacuum stability without the Landau pole, and $\lambda_\Phi  = 10^{-3}$.
The value of $\lambda_1=\lambda_2=\lambda_H$ at $\mu=v_\Phi$ has been evaluated by 
  extrapolating the SM Higgs quartic coupling with $M_h=125$ GeV from the electroweak scale to $v_\Phi$. 
For this parameter choice,
 the $Z'$ boson and the right-handed neutrinos have the masses of the same order of magnitude as
 $M_{Z'}=3.4$ $(34)$ TeV and $M_N=2.0$  $(20)$ TeV for $v_\Phi=10$ $(100)$ TeV,
 while the $B-L$ Higgs boson mass is calculated as $M_\phi=0.23$ $(2.3)$ TeV. 
As is well-known, $M_\phi \ll M_{Z'}$ is a typical prediction of the CW mechanism.
The masses of the heavy Higgs bosons are roughly 1\,TeV for both $v_\Phi=10$\,TeV and 100\,TeV.

In order for the bosonic seesaw mechanism to work,
 we have assumed the hierarchy among the quartic couplings as 
 $\lambda_{H1\Phi} \ll \lambda_{\rm mix} \ll \lambda_{H2\Phi}$ at the scale $\mu=v_\Phi$.
One may think it unnatural to introduce this large hierarchy by hand.
However, we find from Fig.~\ref{running} that the large hierarchy 
  between $\lambda_{H1\Phi}$ and $\lambda_{H2\Phi}$ tends to disappear toward high energies.
This is because the beta functions of the small couplings
 $\beta_{\lambda_{H1\Phi}}$ and $\beta_{\lambda_{H2\Phi}}$  are not simply proportional to themselves,
 but include terms given by other sizable couplings,
 such as $\lambda_3 \lambda_{H2\Phi}$ and $g_{B-L}^4$. 
This behavior of reducing the large hierarchy in the renormalization group evolutions  
   is independent of the choice of the boundary conditions for $g_{B-L}$, $\lambda_3$, $\lambda_4$  and $\lambda_\Phi$. 
Therefore, Fig.~\ref{running} indicates that once our model is defined at some high energy, say, the Planck scale, 
   the large hierarchy among the quartic couplings, which is crucial for the bosonic seesaw mechanism to work, 
   is naturally achieved from a mild hierarchy at the high energy.

\begin{table}[t]
\begin{center}
\begin{tabular}{|c|cc|}\hline
 & $SU(3)_c \otimes SU(2)_L \otimes U(1)_Y$ & $U(1)_{B-L}$ \\
\hline \hline
$S_{L,R}$ & (1, 1, 0) & $x$\\
$S'_{L,R}$ & (1, 1, 0) & $x-2$\\
$D_{L,R}$ & (1, 2, 1/2) & $x$\\
$D'_{L,R}$ & (1, 2, 1/2) & $x+2$\\
\hline
\end{tabular}
\end{center}
\caption{Additional vector-like fermions. $x$ is a real number.}
\label{table:3}
\end{table}

We see in Fig.~\ref{running} that $\lambda_{\rm mix}$ is almost unchanged.
This is because $\beta_{\lambda_{\rm mix}}$ is proportional to $\lambda_{\rm mix}$, 
  which is very small.
Hence, the hierarchy between $\lambda_{\rm mix}$ and the other couplings gets enlarged at high energies.
To avoid this situation and make our model more natural, 
   one may introduce additional vector-like fermions listed in Table \ref{table:3}, for example.
(As another possibility, one may think that some symmetry forbids the $\lambda_{\rm mix}$ term
  and it is generated via a small breaking.)
Although $x$ is an arbitrary real number, we assume $x\neq 1$ to distinguish the new fermions from the SM leptons.
These fermions have Yukawa couplings as
\begin{eqnarray}
	-{\mathcal L}_V &=&  Y_{SS} \overline{S_L} \Phi S'_R + Y_{SD} \overline{S'_R} H_2^\dagger D'_L
		+ Y_{DD} \overline{D'_L} \Phi D_R \nonumber \\
		&& + Y_{DS} \overline{D_R} H_1 S_L  + Y'_{SS} \overline{S_R} \Phi S'_L 
		+ Y'_{SD} \overline{S'_L} H_2^\dagger D'_R \nonumber \\
		&& + Y'_{DD} \overline{D'_R} \Phi D_L + Y'_{DS} \overline{D_L} H_1 S_R
					+ h.c.,
\end{eqnarray}
 so that $\beta_{\lambda_{\rm mix}}$ includes
 terms of $Y_{SS} Y_{SD} Y_{DD} Y_{DS}$ and $Y'_{SS} Y'_{SD} Y'_{DD} Y'_{DS}$,
 which are not proportional to $\lambda_{\rm mix}$.
%
%
%
%
Accordingly, the minimization condition of $V_\Phi$ is modified to
\begin{eqnarray}
	\lambda_\Phi &\simeq& \frac{11}{6 \pi^2}
		\left[ 
		- \frac{1}{8}\left( Y_{SS}^4 + Y_{SS}^{\prime 4} + 2Y_{DD}^4 + 2Y_{DD}^{\prime 4} \right)
		\right. \nonumber \\
		&& \left. \frac{}{} + 6 g_{B-L}^4 - {\rm tr} Y_M^4 \right].
\label{CW_relation2}
\end{eqnarray}
From the conditions $\lambda_\Phi>0$ and $g_{B-L}<0.2$,
 the additional Yukawa contribution should satisfy
 $Y_{SS}^4 + Y_{SS}^{\prime 4} + 2Y_{DD}^4 + 2Y_{DD}^{\prime 4} \lesssim 3\times(0.4)^4$.
Note that the vector-like fermions masses are dominantly generated by $v_\Phi$,
 and they are sufficiently heavy to avoid the current experimental bounds.

\begin{figure}[t]
\begin{center}
          \includegraphics[clip, height=6cm]{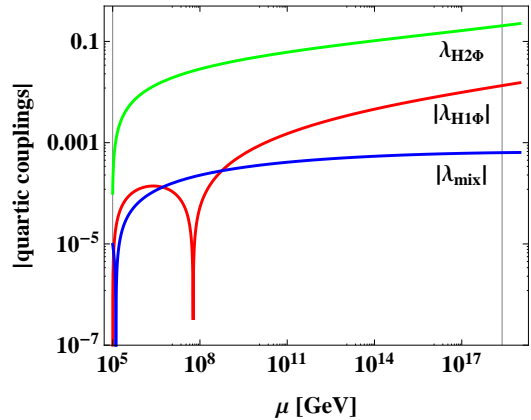}
\end{center}
\caption{Runnings of quartic couplings for $v_\Phi=100$\,TeV
 with additional vector-like fermions.
The vertical axis shows absolute values of quartic couplings.
The input parameters are the same as before.}
\label{running2}
\end{figure}

Fig.~\ref{running2} shows the runnings of the quartic couplings for $v_\Phi=100$ TeV
  with the additional vector-like fermions. 
The input parameters are the same as before,
  while we have taken the Yukawa couplings as
  $Y_{SS}=Y_{SD}=Y_{DD}=Y_{DS}=0.2$ and $Y'_{SS}=Y'_{SD}=Y'_{DD}=Y'_{DS}=0.1$ at $\mu=v_\Phi$, 
   for simplicity. 
Toward high energies, $|\lambda_{\rm mix}|$ becomes larger,
   and the hierarchy with the other couplings becomes mild. 
We can see that $\lambda_{H1\Phi}$ is negative below $\mu\simeq10^8$ GeV,\footnote{
Although $\lambda_{H1\Phi}$ and $\lambda_{\rm mix}$ become negative,
 their values can be much smaller than the self couplings ($\lambda_1$ and $\lambda_2$).
Thus, the vacuum is stable in our model.}
  because the contributions of additional Yukawa couplings to $\beta_{\lambda_{H1\Phi}}$ are effective
  below $\mu\simeq10^8$ GeV.
Above the scale, the contribution of $U(1)_{B-L}$ couplings becomes effective,
  and then $\lambda_{H1\Phi}$ becomes positive.
As a result, the large hierarchy at the $U(1)_{B-L}$ symmetry breaking scale can be realized 
   with a mild hierarchy at some high energy.  
We expect that a ultraviolet complete theory, which provides the origin of 
  the classical conformal invariance, takes place at the high energy.

\section*{Acknowledgment} \label{Acknowledgement}
N.O. would like to thank the Particle Physics Theory Group of Shimane University
  for hospitality during his visit.
This work is partially supported by Scientific Grants
  by the Ministry of Education, Culture, Sports, Science and Technology (Nos. 24540272, 26247038, and 15H01037)
  and the United States Department of Energy (DE-SC 0013680).
The work of Y.Y. is supported
  by Research Fellowships of the Japan Society for the Promotion of Science for Young Scientists
  (Grants No. 26$\cdot$2428).


\end{document}